**Two-dimensional borophene: In-plane hyperbolic polaritons in the visible spectral range**

*Yaser Abdi[1,2]\*, Masoud Taleb[2], Soheil Hajibaba[1], Mohsen Moayedi[1], Nahid Talebi[2]\**

1. Department of Physics, University of Tehran, 1439955961 Tehran, Iran

2. Institute of Experimental and Applied Physics, Kiel University, 24118 Kiel, Germany

   \* E-Mail: y.abdi@ut.ac.ir , talebi@physik.uni-kiel.de

**Abstract -** Two-dimensional metals, such as graphene, have been extensively explored, with graphene exhibiting a metallic response limited to the infrared spectral range. Extending the electron mobility in two-dimensional metals to achieve plasmonic behaviors in the visible range requires innovative synthesis procedures. In this study, we successfully realized the $\chi_3$ phase of borophene using chemical vapor deposition. Leveraging first-principle density-functional theory alongside advanced deep-subwavelength cathodoluminescence spectroscopy, we reveal the extreme anisotropic response of this material in the visible range, transitioning from hyperbolic polaritonic to an elliptic wavefront. Our calculations substantiate the experimental findings, positioning borophene as a candidate for the in-plane hyperbolic response in the visible range. These results open an avenue for optoelectronic applications in the visible spectrum, particularly through the incorporation of borophene into hybrid metallic-semiconducting heterostructures.

**Keywords:** hyperbolic polaritons, visible range, anisotropic, borophene, two-dimensional

## Introduction

Extreme confinement of light and tailoring its coupling to matter are at the heart of advanced solid-state quantum technologies. Polaritons are quasiparticles formed by the strong interaction of photons with a matter excitation. In particular, various forms of polaritons in two-dimensional (2D) materials are attractive candidates for further tailoring of strong light-matter interactions, due to the reduced attenuation and screening effects[1,2]. Moreover, 2D materials offer a great versatility in forming hybrid material platforms, favoring quasi-particle interactions[3-5] and enhanced light confinement due to their reduced dimensionality[6].

In addition to tailoring the dimensionality, reducing the symmetry of the crystal structure of 2D materials leads to even more functionalities[7]. Extreme anisotropic responses that lead to hyperbolic isofrequency surfaces[8,9] and directional propagation[10] enable a stronger coupling to localized emitters[11,12], enhancing the Purcell factor of quantum emitters[13,14], and achieving superlens functionalities[15]. Several metamaterial concepts have been suggested and realized[16,17] to enable the hyperbolic isofrequency surfaces in specific frequency ranges. Nevertheless, and in parallel to the metamaterials research, a large number of natural materials have recently emerged that exhibit hyperbolic polaritonic properties. Bulk and surface phonon polaritons in different kinds of materials, including uniaxial crystals such as hexagonal boron nitride[18-21] as well as biaxial crystals such as $\alpha$-MoO$_3$[22,23], have been demonstrated. It has been shown that hyperbolic polaritons in $\alpha$-MoO$_3$ films can be tuned by doping and an applied gate voltage[24-26]. Shear polaritons have been as well reported in the CdWO$_4$ monolithic crystal within the mid infrared spectralrange[27]. Phonon-polaritons are naturally excited in the far infrared to the mid infrared spectral range, not accessible to the present spectral range of single-photon emitters and quantum technological applications. Therefore, hyperbolic materials in the visible range[28] appear as better candidates for quantum-optical applications, such as coupling to defect centers in 2D materials and vacancy centers in diamond[29].

A few materials including tetradymites have been recently explored due to their highly non-symmetric uniaxial crystal structure and have been shown to enable hyperbolic surface polaritons within the visible range[30-32].Biaxial and in-plane hyperbolic features in the visible range in a natural material have been recently reported in thin films of MoOCl$_2$[33]. Here, we demonstrate an alternative two-dimensional material, borophene, which exhibits the excitation of in-plane hyperbolic polaritons, similar to MoOCl$_2$.

Borophene has recently attracted significant attention, especially after the first synthesis of this material on a bulk metal[34,35]. It has been demonstrated as a prototype of synthetic 2D materials, without having a van der Waals bulk counterpart[36]. Along with the recent advancements in synthesis techniques[37], the stability of the borophene sheets has been explored as well in different environmental conditions[38,39]. Borophene's stability and its metallic response makes it suitable for a wide range of applications including energy storage[40,41], catalysts[42], flexible electronics[43], gas sensors[44], and biomedical applications[37]. Despite its potential, producing high-quality borophene at a large scale remains a challenge. Furthermore, integrating borophene into existing technologies and devices requires further research and development.

It has been shown theoretically in 1990 that several classes of boron sheets should support free conduction electrons and behave as a metal[45]. More recently, borophene was suggested to enable a plasmonic response within the visible range, thanks to the high mobility ($\mu$) of the conduction electrons in this material, being at the order of $\mu = 28.4 \times 10^5$ cm$^2$ V$^{-1}$ s$^{-1}$ [46,47]. Given the remarkable application of visible-range plasmonic properties in nano and 2D materials[48-50], exploring the plasmonic behavior of borophene is crucial. However, the optical response of free-standing borophene sheets, has not been experimentally explored, because the previous synthesis methods have been used to deposit a borophene sheet on a pre-existing bulk metal such as silver.

Here we realize borophene films in $\chi_3$ phase on different substrate platforms using chemical vapour deposition (CVD)[41,51]. For investigating the optical properties of borophene, it is particularly favorable to realize free-standing borophene. We first synthesize borophene on holey carbon and gold grids, and apply cathodoluminescence (CL) spectroscopy[52,53] to explore the optical properties of borophene sheets. We further study its crystal structure using analytical transmission electron microscopy techniques as well as Raman spectroscopy. Using density-functional theory (DFT) based on Quantum ESPRESSO open-source package, we further calculate the permittivity of single-layer borophene sheets. We further demonstrate, both theoretically and experimentally, that borophene supports extremely confined in-plane hyperbolic polaritons over a broad wavelength range of 521 nm < λ < 1100 nm, due to the optical axis of this highly anisotropic crystal being oblique with respect to the sheet's normal. Using CL spectroscopy, we further confirm the directionality of optical excitations in $\chi_3$ borophene flakes. Our explorations confirm that borophene is not only a 2D metal within the visible range, but also exhibits an extremely anisotropic response and highly directional hyperbolic polaritons within the visible spectral range. Therefore, we suggest borophene in applications that enable strong interactions with highly confined quantum emitters and tailoring the radiation from such emitters.

**Results**

**In-plane Polaritons in Borophene**

Our synthesis method, which enables the realization of $\chi_3$ borophene on arbitrary substrates, is explained in the Methods section. Here, we first discuss the key features expected from the extremely asymmetric crystal structure of single-layer borophene sheets. The $\chi_3$ borophene structure consists of a mixture of triangular and hexagonal motifs, with two in-plane principal axes oriented along the directions indicated by the x and y arrows, which are perpendicular to each other (see Fig. 1b). The third principal axis along the *z*-direction is normal to the in-plane principal axes. $\chi_3$ consists of atomic ribbons oriented along the *y* direction, resembling a macroscopic grating-like structure of parallel graphene ribbons with hyperbolic properties[54]. This unique crystal structure results in asymmetric electronic and optical properties. Additionally, due to its unique crystal structure, borophene features asymmetric Dirac cones. Charge transfer along the triangular motif differs from transport in the direction perpendicular to the triangular motif, contributing to borophene's anisotropic electronic behavior. The calculated band structure of $\chi_3$ borophene, depicted in Supplementary Fig. 1a, follows a G-M-X-G-N-Y-G path in the reciprocal space of the Brillouin Zone. This electronic band structure reveals the distinct anisotropy of $\chi_3$ borophene. In certain directions, such as X-G, the electron energy bands cross the Fermi level, indicating metallic behavior. Instead, in the X-M direction, the electron bands near the Fermi level are relatively flat, suggesting more localized electronic states and directional variation in electronic properties. Furthermore, $\chi_3$ borophene has a nonzero density of states (DOS) at the Fermi level (Supplementary Fig. 1b). Therefore, $\chi_3$ borophene exhibits metallic behavior with strong anisotropy, and its electrical conductivity is confined along the atomic strips parallel to the *y* direction. The details of our DFT calculation are provided in Methods Section.

The calculated permittivity of the $\chi_3$ borophene sheets indeed demonstrates the extreme anisotropic response of the material (Fig. 1a): whereas the permittivity along the principal axis

*x* features a dielectric response, the response along the *y*-axis is Drude metallic with the plasma wavelength located at $\lambda_\text{P} = 521$ nm ($E_\text{p} = 2.38$ eV).

The Drude-like material's permittivity along the *y* direction next to the dielectric response along the *x* direction, comprise two domains for the optical response of the material. Within the wavelength range of $\lambda > \lambda_\text{P}$, the in-plane isofrequency surfaces of the material demonstrate type II hyperbolic features, whereas for the wavelengths $\lambda < \lambda_\text{P}$, the isofrequency surfaces have elliptical forms (Fig. 1c). Note that the wavenumbers are complex-valued, demonstrating a nonvanishing attenuation of the optical waves confined to the borophene sheets with their propagation direction lying in the *x-y* plane.

The dispersion of optical plane waves propagating in borophene is derived by calculating the eigen values of the Maxwell's equation at each given frequency as $\vec{k} \times \vec{k} \times \vec{E} + k_0^2 \hat{\varepsilon}_r : \vec{E} = 0$, where $\vec{k} = k_x \hat{x} + k_y \hat{y} + k_z \hat{z}$ is the wave vector, $\hat{\varepsilon}_r$ is the permittivity tensor composed of only diagonal terms, $k_0$ is the free-space wavenumber, and $\vec{E}$ is the electric field. For waves traveling along the *x-y* plane, we set $k_z = 0$, and obtain $\left(k_x^2 / \varepsilon_{r,xx}\right) + \left(k_y^2 / \varepsilon_{r,yy}\right) = k_0^2$, that is an elliptical response for $\varepsilon_{r,xx} > 0$ and $\varepsilon_{r,yy} > 0$ and a hyperbolic response for $\varepsilon_{r,xx} > 0$ and $\varepsilon_{r,yy} < 0$.

We further examine the possibility of the propagation of in-plane polaritons in the borophene sheet. For highly anisotropic and particularly hyperbolic materials, several forms of bulk[18], surface[55], edge[31], shear[56], and ghost[57] polaritons have already been demonstrated in the literature. For an atomically thin material like borophene, in-plane hyperbolic polaritons can be better captured by considering a negligible thickness for the material. Therefore, we translate the permittivity along *x* and *y* components into the in-plane conductivity tensor $\hat{\sigma}$, as $\hat{\sigma} = -i\omega d\varepsilon_0 \left(\hat{\varepsilon}_r - 1\right)$, where $\hat{\varepsilon}_r$ is the in-plane relative permittivity tensor and $d$ is the thickness of the borophene film[58]. As shown in the Supporting Information Note 2, the spatial distributions associated with the field components of in-plane polaritons, for the borophene sheet located at $z = 0$ plane is constructed as $\psi\left(\vec{r}_\|, z\right) = \psi_0 \exp\left(-\kappa_z |z|\right) \exp\left(i\vec{k}_\| \cdot \vec{r}_\|\right)$, with $\kappa_z^2 = k_\|^2 - k_0^2$ and $\vec{k}_\| = k_\| \cos\varphi \hat{x} + k_\| \sin\varphi \hat{y}$ and $\vec{r}_\| = x\hat{x} + y\hat{y}$ (see the inset of Fig. 1d). Applying the boundary conditions, one obtains the following results:

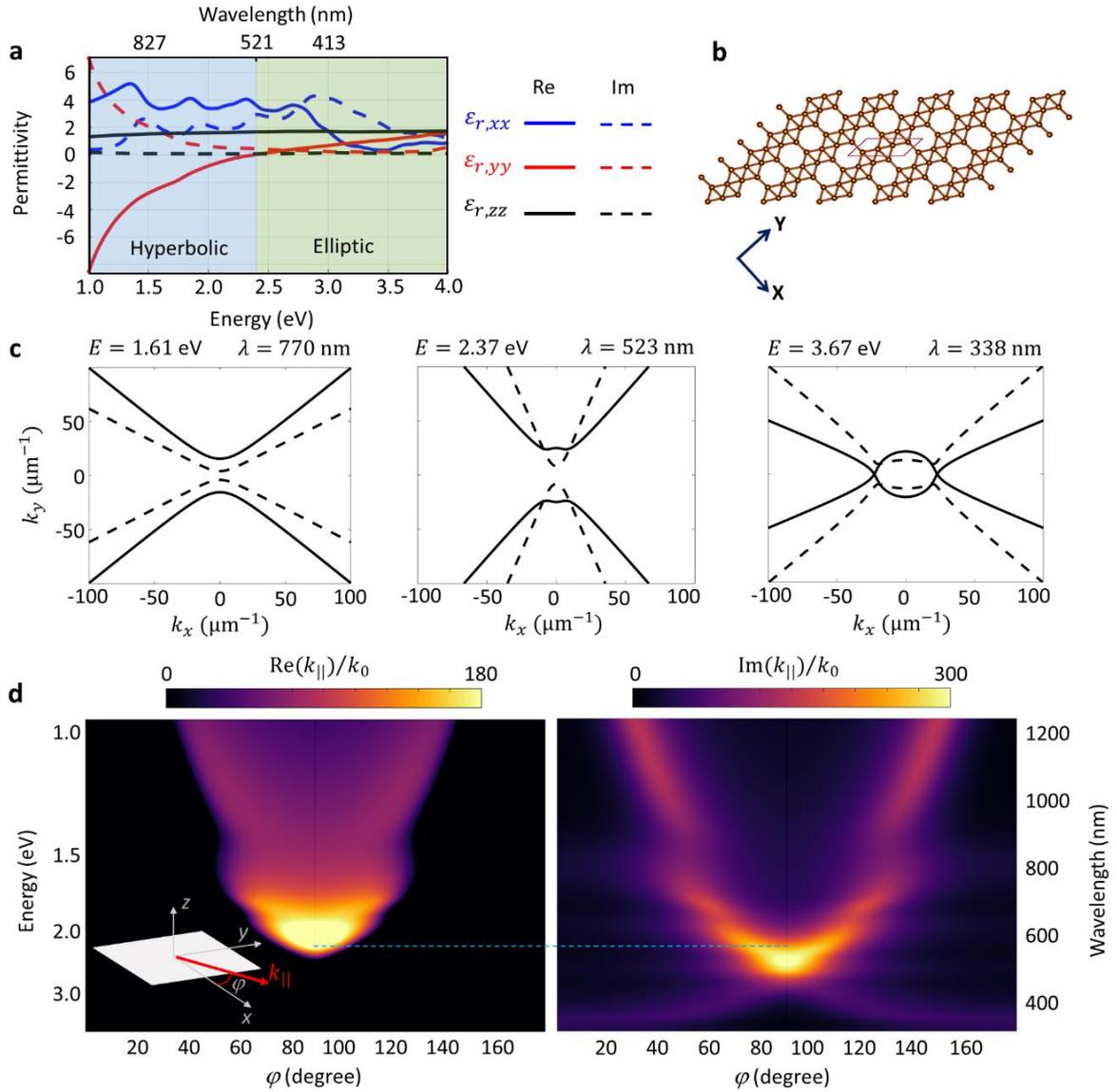

**Fig. 1| Biaxial hyperbolic electromagnetic responses of a Borophene flake.** (a) Real and imaginary parts of the dielectric functions along the three principal axes. Shaded regions demonstrate the photon energy ranges where the optical responses show anisotropic hyperbolic and Elliptic characteristics. (b) Schematic of the crystalline Borophene monolayer in the *x-y* plane. (c) In-plane (*x-y*) iso-frequency surfaces at three depicted energies (wavelengths), showing the transition from in-plane hyperbolic to elliptic responses. Solid and dashed lines denote $\mathrm{Re}\{k_y(k_x)\}$ and $\mathrm{Im}\{k_y(k_x)\}$ respectively. (d) Propagation constant of the in-plane polaritons versus the azimuthal angle with respect to the principal axis *x*. (Left) Phase constant and (Right) attenuation constant. Blue dashed line is inserted as a guide for distinguishing the phase and attenuation constants associated with $\lambda = 600$ nm and $\varphi = 90°$.

$$\kappa_z = \frac{k_0^2}{4i\omega\left(\sigma_{yy}\sin^2\varphi + \sigma_{xx}\cos^2\varphi\right)} \cdot \left\{ -\left(\frac{4}{\mu_0} + \frac{\sigma_{xx}\sigma_{yy}}{\varepsilon_0}\right) \right.$$
$$\left. \pm \sqrt{\left(\frac{4}{\mu_0} + \frac{\sigma_{xx}\sigma_{yy}}{\varepsilon_0}\right)^2 - 16c^2\left(\sigma_{yy}\sin^2\varphi + \sigma_{xx}\cos^2\varphi\right)\left(\sigma_{xx}\sin^2\varphi + \sigma_{yy}\cos^2\varphi\right)} \right\}, \quad (1)$$

where $k_0 = \omega/c$ is the free-space wavenumber, $\varepsilon_0$ and $\mu_0$ are the free-space permittivity and permeability, respectively, $\omega$ is the angular frequency and $c$ is the speed of light in vacuum. $\varphi$ is the angle of the wave vector of the in-plane polaritons with respect to the principal axis x.

Using equation (1), the propagation constant of in-plane polaritons is obtained as $k_\parallel = \sqrt{\kappa_z^2 + k_0^2}$. The real part of $k_\parallel$, which represents the phase constant of polaritons in borophene, has a non-zero value for only a specific range of angular distributions. This fact confirms the directional propagation of hyperbolic polaritons. Particularly along the x-axis the borophene sheet cannot sustain any form of interface modes, as expected. In contrast, the largest phase constant occurs for polaritons propagating along the y-axis at the wavelength of 600 nm (E = 2.06 eV), where the permittivity of the material along the y-axis is $\epsilon_{r\,yy} \cong -1$. At this wavelength, the effective wavelength of in-plane polaritons is 300 times smaller than the free-space wavelength. At longer wavelengths, in-plane polaritons demonstrate an optimum directional propagation at an angle different from $\varphi = 90°$, as shown in Fig. 1d.

**Borophene Synthesis and Structural Characterization**

Borophene sheets grown by the aluminum-assisted CVD approach exhibit varying thicknesses and shapes (See Supplementary Fig. 2). An in-depth investigation of the thickness, crystal structure, and chemical stability of borophene grown by this aluminum-assisted CVD approach has been conducted previously[45]. Our transferring technique, detailed in the Methods section, effectively transfers CVD-grown borophenes onto desired substrates, such as gold mesh and carbon-coated copper grid which are used here for transmission electron microscopy and cathodoluminescence characterizations (see the secondary electron SEM image, Fig. 2a and b). The Raman spectra of borophene further reveal its structural characteristics. The peaks around 180 cm$^{-1}$, 255 cm$^{-1}$, 305 cm$^{-1}$, 415 cm$^{-1}$, and 445 cm$^{-1}$ (Fig. 2c) are in good agreement with the B$_g^2$, B$_g^2$ (X'), A$_u$(Y), B$_g^1$, and B$_g^1$ (Y) mods, respectively, which were theoretically predicted for the χ$_3$ phase[59]. The peak intensities of the Raman spectra (Fig. 2c) are comparable with the reported Raman spectra[60]. The low Raman peak intensity of borophene is due to the light mass of boron atoms, which results in a smaller Raman scattering cross-section compared to graphene and silicene.

The high-resolution transmission electron microscopy (HRTEM) image of the borophene sheets shows the parallel lattice strips with an interstrip distance of 4.4 and 2.1 angstroms (Fig. 2d, 2e), which is in excellent agreement with the distance between parallel strip-like regions of high atomic concentration and the atomic distance in the $\chi_3$ phase, respectively. Furthermore, selected area electron diffraction (SAED) of the structures (Fig. 2f) exhibit an excellent consistency with the primary reciprocal lattice of $\chi_3$ borophene. The Miller indices corresponding to the observed diffraction pattern are determined by comparing the theoretical values of the *d*–spacing with a standard deviation of 0.005.

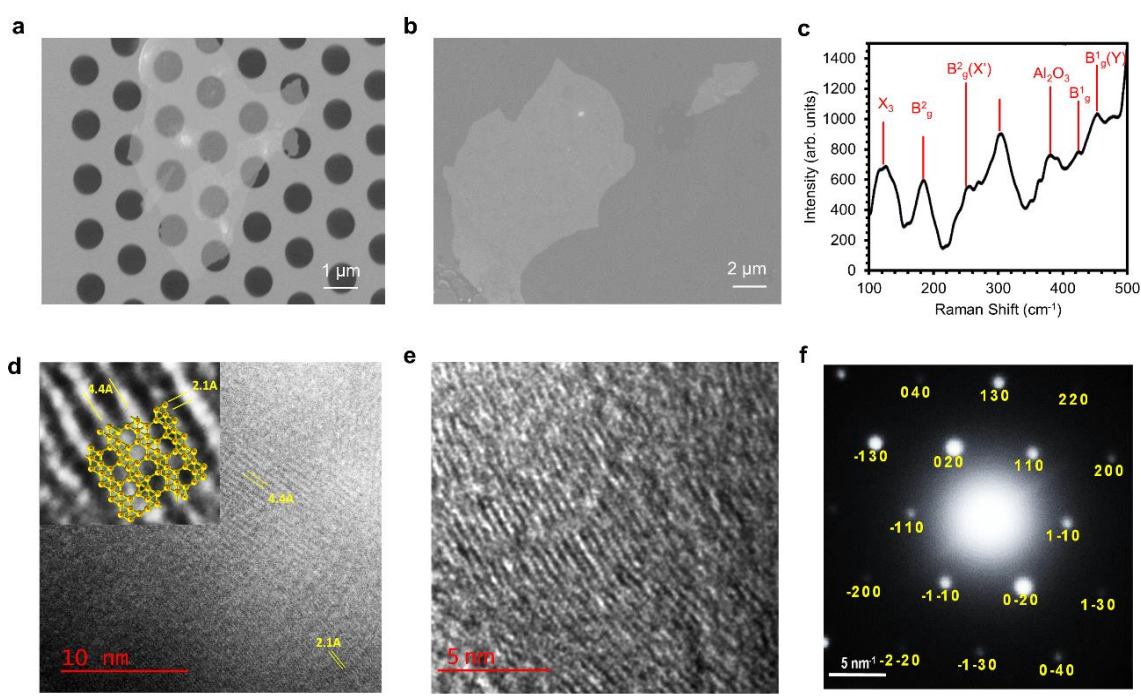

**Fig. 2| Morphological and structural analyses.** (a) SEM image of the grown borophene sheets on a gold mesh grid. (b) SEM image of borophene on a carbon-coated copper grid. (c) Far-field Raman spectrum of borophene sheets. (d, e) HRTEM image of borophene sheets showing the parallel lattice strips with interstrip distance of 4.4 and 2.1 angstroms which is in excellent agreement with the distance between parallel strip like regions of high atomic concentration and atomic distance in the χ3 sheet, respectively. (f) SAED of borophene which are well-matched with the crystal structure of the χ3 phase.

The structural analysis results reveal the unique crystalline structure of borophene, setting it apart from other extensively studied 2D materials. Unlike graphene, which possesses a single-phase hexagonal lattice—a symmetric crystal structure that leads to a symmetric Dirac cone in its band structure—borophene exhibits a mixed arrangement of triangular and hexagonal motifs. This distinctive crystal structure results in asymmetric electronic and optical properties. Furthermore, due to its two-phase crystal structure, borophene features asymmetric Dirac cones. Carrier mobility along the triangular motif differs from that in the direction perpendicular to the triangular motif, contributing to borophene's anisotropic electronic behavior.

**Cathodoluminescence Spectroscopy of In-Plane Hyperbolic Polaritons in Borophene**

To explore the optical response of borophene flakes, we use here CL spectroscopy, which allows us to reveal the spatio-spectral features of propagating[31,61-67] and localized[68-74] polaritons in different bulk and 2D materials. Using CL spectroscopy, photons generated from the interaction of electron beams with samples are analyzed. CL allows for accessing the radiative local density of states, whereas electron energy-loss spectroscopy allows for revealing the full projected local density of states along the electron trajectory[75]. Moreover, CL has several advantages, including having a higher spectral resolution compared to electron energy-loss spectroscopy, allowing for performing polarimetry, and exploring the photon statistics of excitations[76]. Moreover, CL is readily employable in SEMs where the ample space within the chamber allows for multi-sequential experiments and performing spectral interferometry[60].

In our experimental scheme, the photons generated from the interaction between the electron beam and borophene flakes are collected using a parabolic mirror and directed toward our analyzing path (Fig. 3a). We perform spectroscopy, angle-resolved mapping, and polarimetry, to unravel the characteristics of localized hyperbolic polaritons in the synthesized borophene flakes.

We first investigate the cathodoluminescence (CL) response of a borophene sheet transferred onto a copper grid coated with a holey carbon film (see the secondary electron SEM image in Fig. 3b). This substrate facilitates the formation of localized hyperbolic polaritons due to the partial reflection of propagating hyperbolic polaritons from the carbon boundaries. At $\lambda$ = 485 nm, the spatial distribution of the maximum CL signal originates from the boundaries and follows the morphology of the holey carbon structure, forming an elliptical propagation pattern (see Fig. 3c, upper panel). At longer wavelengths (λ ≥ 521 nm), in-plane hyperbolic polaritons are excited within the borophene sheet (see the hyperspectral image in Fig. 3c at λ = 650 nm). Due to the directional propagation of hyperbolic polaritons, the spatial distribution of localized polaritons reflected from the boundaries does not fully conform to the boundary morphology. Furthermore, additional fringes appear along the horizontal x-axis but are absent along the vertical y-axis, highlighting the anisotropic nature of the polaritons in borophene.

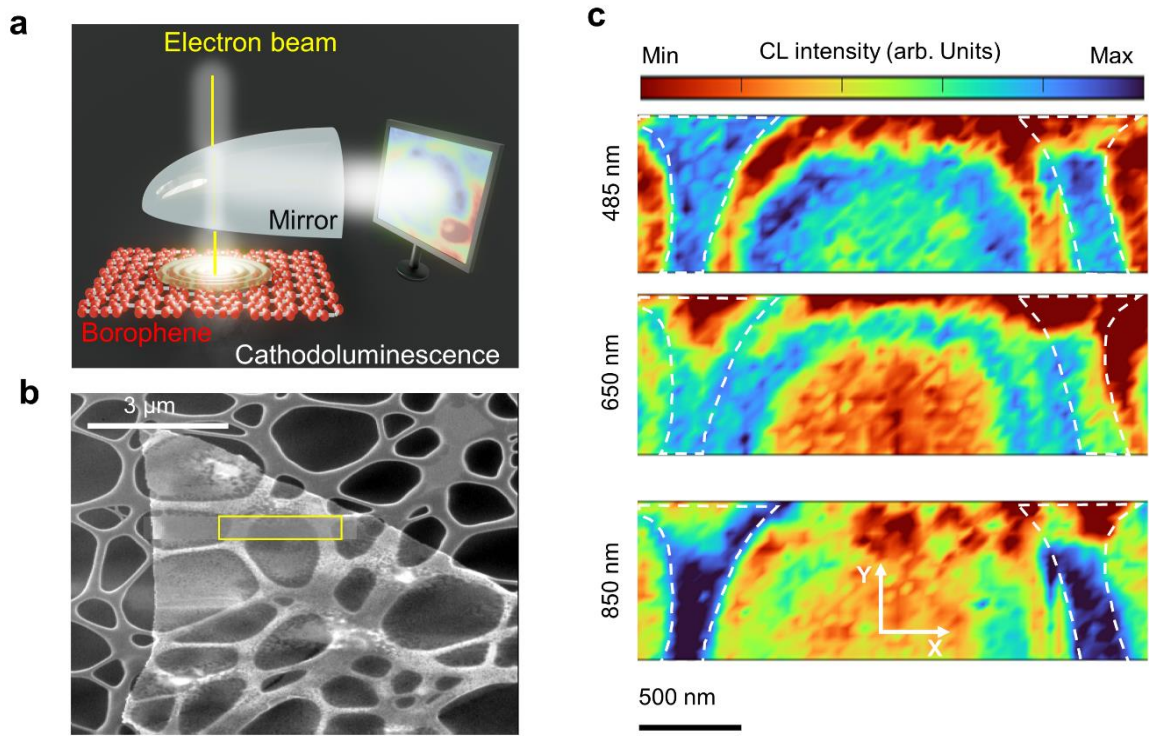

**Fig. 3| Cathodoluminescence of borophene sheet on holey carbon**. (a) Experimental setup consisting of a parabolic mirror to gather radiation emitted from the sample and collimating the radiation toward the analyzing path. (b) Secondary-electron SEM image of the Borophene sheet on holey carbon-coated TEM grid. (c) Hyperspectral images at selected wavelengths taken from region indicated in panel by the yellow box (b). The dashed lines indicate the boundaries of holey carbon.

We next consider a borophene nanoparticle with a simple square shape, that allows us to better reveal the anisotropic characteristics of the hosted polaritons (Fig. 4a). The synthesized nanostructure is positioned on a Si substrate.

In-plane hyperbolic polaritons are expected to be excited at wavelengths longer than 521 nm when the material exhibits a hyperbolic response (Fig. 1a and d). Acquired CL signals at wavelengths longer than $\lambda = 521$ nm indeed have a significantly larger intensity that originates from a stronger CL spontaneous yield, due to the excitation of hyperbolic polaritons. Whereas the spatial profile of the CL intensity at $\lambda = 495$ nm follows the morphology of the boundaries, at $\lambda = 550$ nm the CL intensity exhibits two strong emission cites at only left and right boundaries, originated from the direction propagation of polaritons along the path shown by the arrow (See Fig. 4b, top right panel). This results in a dipole excitation (Fig. 4c), that is further revealed by the angle-resolved polarimetry (Fig. 4e and f). Retrieved Stokes parameter of the far-field signal precisely determine the radiation pattern of a dipolar charge oscillation. More precisely, $S_1 = |E_x|^2 - |E_y|^2$ shows an x-polarized like along the y-axis and ay-polarized light along the x-axis, associated with the dipolar radiation pattern. The reconstructed z-component of the electric field demonstrates a donut-like shape, which is expected due to the transversal nature of the far-field radiation. The azimuthal component of the electric field demonstrates two lobes at $\varphi = 0$ and $\varphi = 180°$, and zero intensities at $\varphi = \pm 90°$. This behavior is attributed to the excitation of charges accumulated at rims along the *y*-axis, that further forms a dipolar resonance.

At longer wavelengths, λ=750 nm and 850 nm, the in-plane polaritons exhibit smaller fringe spacing, which enhances the visibility of their anisotropic propagation. Moreover, by representing the CL spectra versus position along the path shown by dashed blue line in Fig 4a (See Fig. 4d), the excitation of quadrupolar modes at the wavelength of 820 nm is revealed. The wavelength-dependent position of the maximum of the CL intensity depicts a decrease in the fringe spacing of polariton propagation as well. Such anisotropic propagation has been previously seen in disk-like materials with in-plane hyperbolic polaritons[22].

To better confirm that the CL signal originates from borophene and not the Si substrate, we analyze the CL spectra versus various kinetic energies for the electron beam (Fig. 4g). The CL signal exhibits a maximum at the kinetic energy of 10 keV with distinguished peaks at 520 nm and 820 nm, allocated to the dipolar and quadrupolar resonances respectively. For larger kinetic energy of the electron beam, the beam penetrates deeply into the Si substrate and the CL yield from borophene decreases. We observe a rather broad spectral feature when the electron interacts with the borophene sheet. The absence of a sharp resonance is expected and is attributed to the presence of a large photonic density of state supported by the hyperbolic materials.

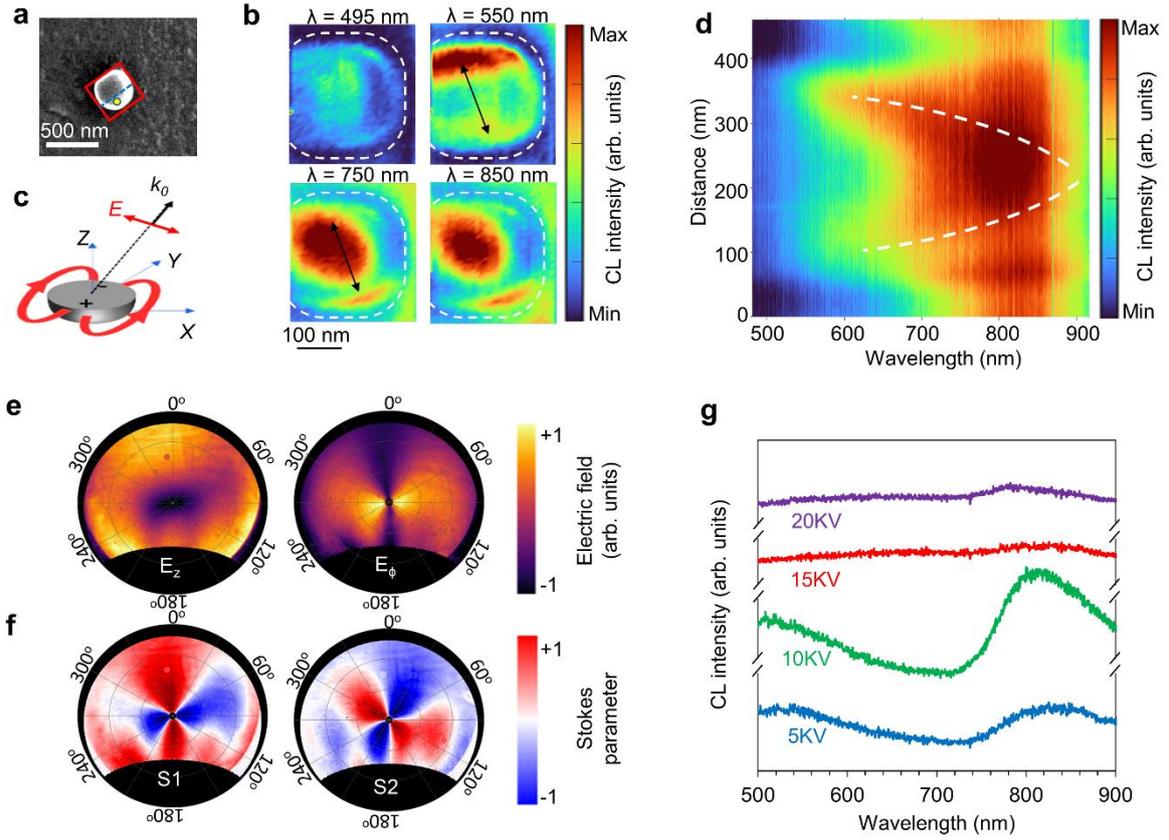

**Fig. 4| Cathodoluminescence of Borophene flakes on Si. (a)** Secondary-electron scanning Electron Microscopy image of a borophene flake **(b)** Hyperspectral images of the selected area marked by the red square in panel (a) at selected wavelengths (depicted on each image), illustrating the localization of hyperbolic polaritons either at the rim along the *y*-axis or at the centre. **(c)** Excitation of a dipolar-like optical mode upon electron impact and far-field CL signal originated from this excitation. **(d)** Distance-wavelength spectral distribution of CL light at the positions marked by the dashed cyan line in panel (a). **(e)** Retrieved spatial profile of the $E_z$ and $E_\varphi$ components of the far-field electrical field distribution. **(f)** $S_1$ and $S_2$ Stokes parameters measured at $E$ = 2.25 eV ($\lambda \approx 550$ nm) for the electron impact position shown in panel (a). **(g)** CL spectra acquired at same selected electron impact position marked in the SEM image of panel (a) at depicted electron beam acceleration voltages with a beam current of 6 nA.

Therefore, in-plane polaritons in borophene manifest a wavelength-dependent propagation direction. In addition, they propagate at shorter lengths compared to plasmon polaritons in typical noble materials such as gold. The propagation length of in-plane polaritons in borophene can be inferred from the calculated attenuation constant as $L(\lambda) = \alpha(\lambda)^{-1}$. For polaritons propagating at the wavelength of $\lambda = 600$ nm ($E = 2.07$ eV), this results in only 450 nm propagation length. At longer wavelengths though, the attenuation constant decreases, becoming 900 nm at = 1100 nm ($E = 1.13$ eV). Due to the large attenuation constant of borophene in-plane polaritons within the given wavelengths, and additionally their directional propagation, hyperspectral images can demonstrate interference fringes only at distances near an edge. Such interference fringes generally form due to the interference between different pathways, such as transition radiation interfering with polaritons scattered from the edges[52], or the interference between forwardly propagating waves and waves reflected from the edges of the flakes[77]. The strong absorption rate of borophene lead to improvements in the efficiency of optoelectronic devices such as photodetectors[78]. Nevertheless, directional propagation of the polaritons can be inferred from CL hyperspectral images acquired by scanning the structure along different crystallographic axes perpendicular to an edge of the flake, as we show below.

The direct synthesis of borophene on $SiO_2$/Si substrates yields higher-quality flakes compared to previous samples grown on Si. Moreover, depending on the growth parameters, the flakes can exhibit a multilayer structure. We estimate their thickness to be approximately 10 nm (measured by tilting the sample to 90° in the SEM and imaging the structure via the in-lens detector with approximately 1 nm resolution). These flakes were transferred from $SiO_2$/Si onto a holey carbon substrate for CL spectroscopy (Fig. 5a).

Hyperspectral images acquired along the direction perpendicular to the edge of a pristine flake reveal wavelength-dependent minima in the CL intensity, which appear parallel to the edge (Fig. 5b, right). At a wavelength of 600 nm ± 10 nm, this minimum is positioned 90 nm away from the edge, while at longer wavelengths, it shifts further away. At 600 nm, in-plane polaritons in a borophene film with a thickness of 10 nm exhibit an effective wavelength of 180 nm and propagate along the y-axis. This value closely matches the distance between the first observed intensity maximum in the CL hyperspectral image and the edge of the flake. Consequently, the direction normal to the left edge of the flake is attributed to the crystalline y-axis.

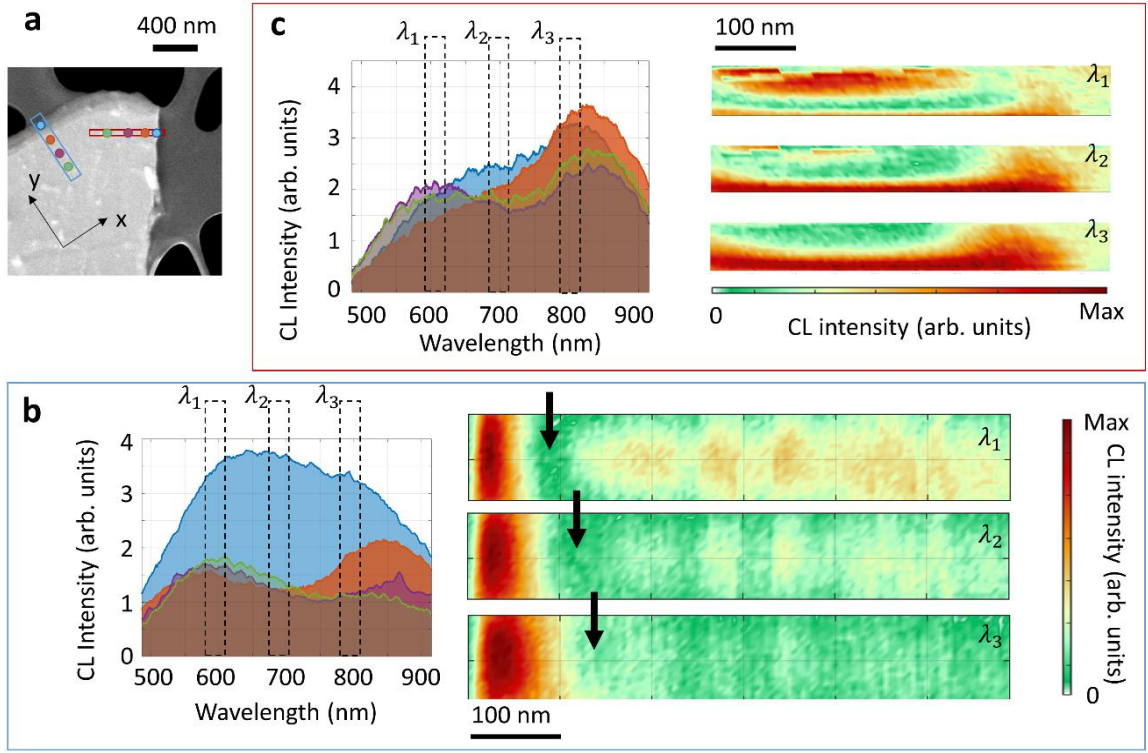

**Fig. 5| Anisotropic polariton propagation in borophene. (a)** Dark-field secondary electron image of a single-crystalline multilayer borophene flake, transferred onto a holey-carbon substrates. **(b)** (Left) CL spectra of the in-plane hyperbolic polaritons at the electron impact positions marked by colored circles in (a) in the region marked by the blue box, and (Right) CL hyperspectral images at depicted wavelengths. **(c)** (Left) CL spectra of the in-plane hyperbolic polaritons at the electron impact positions marked by colored circles in (a) in the region marked by the red box, and (Right) CL hyperspectral images at depicted wavelengths.

CL spectra acquired at the edge of the flake sustain a broad spectral feature, covering the wavelength range from 530 nm to 900 nm (blue area in Fig. 5b and c). At the distance of 100 nm away from the edge, the CL spectrum exhibits two peaks centered at 600 nm and 850 nm, whereas at distances further away from this edge, the spectrum shows only a single resonance at 600 nm. This is due to the fact that only in-plane polaritons propagating along the y-axis can be excited at this wavelength (Fig. 1d).

An electron beam traversing the borophene flake can excite the in-plane hyperbolic polaritons. The polaritons sustaining the wavelength of 600 nm propagate along the y-axis towards the edge of the flake, are further scattered from the edge, and cause radiation in the far-field. However, electron beams generate transition radiation as well, interfering with in-plane polaritons scattered from the edge. This interference phenomenon becomes constructive whenever $\delta = \lambda_{\text{eff}} = 2\pi/\beta$, where $\delta$ is the distance between the maxima of the interference fringes and $\lambda_{\text{eff}} = 2\pi/\beta$ is the effective polariton wavelength.

Therefore, the dominating polaritons are those propagating along the y-axis. In fact, polaritons propagating along a direction other than the y-axis show a large attenuation constant and are largely evanescent, leading to the fast diminishing of the peak centering around 800 nm even at distances as short as 200 nm away from the edge. These evanescent polariton waves contribute to the high-intense CL signal observed at the edges of the flake. Moreover, when the borophene flake is scanned along directions perpendicular to another edge of the flake not being oriented along the crystalline x-axis, no interference fringe being parallel to the edge of the flake is observed (Fig. 5c). In the latter case and in contrast, both peaks at 600 nm and 800 nm survive at different impact positions along the scanned direction. This is due to the interaction of the excited polaritons along different directions interacting with the other edges of the flakes, where the latter are positioned near to the electron impact position as well.

In order to better understand how a moving electron interacts with in-plane polaritons, we have performed time-domain calculations using an in-house developed finite-difference time-domain method. The details of this code can be found elsewhere.[79-81] In the simulations conducted here, the mesh size for the simulation domain is set to 1 nm. The electron probe is modeled using a time-dependent current source with a broadening of 1 nm. The simulation domain is terminated using a higher-order absorbing boundary condition. Additionally, the dielectric function of borophene is modeled by a Drude function, supplemented by two critical point functions, as detailed in the Supporting Information Note 4.

We first present the time-domain results, which demonstrate the dynamics of the interaction (Fig. 6a). Shortly after the interaction begins, two main features are observed. First, a homogeneous ring propagates at the speed of light, manifesting a half-cycle wavefront associated with a rapidly annihilated time-dependent dipole oriented along the z-axis. This excitation is referred to as transition radiation. Transition radiation occurs due to the annihilation of the dipole formed by the moving electron and its image within the thin film, which happens as the electron traverses the interface. Given that the material thickness is 5 nm (equivalent to two unit cells of the simulation domain), the time interval between the two transition radiation events is shorter than the wavelength of the radiation, resulting in only a half-cycle emission.

Second, hyperbolic polaritons form at a slightly delayed time, approximately 4 fs after the transition radiation. Interestingly, even in the time domain, the dominant propagation of hyperbolic polaritons along the y-axis is clearly discernible. The in-plane polaritons reach the edge at t = 11.9 fs, contributing to the radiation paths. A clear reflection from the edge is not observed due to the stronger contribution of the radiation.

Fourier-transformed distributions of the in-plane polaritons further confirm the distinct behavior of the material along different crystallographic axes. Unlike the radial wavefronts formed by the interaction of a moving electron with isotropic noble metals such as gold[81], in-plane polaritons in borophene exhibit a more complex wavefront. The hyperbolic nature of the polaritons becomes evident for wavelengths longer than 600 nm (Fig. 6b and c), whereas for shorter wavelengths, the wavefront appears more elliptical.

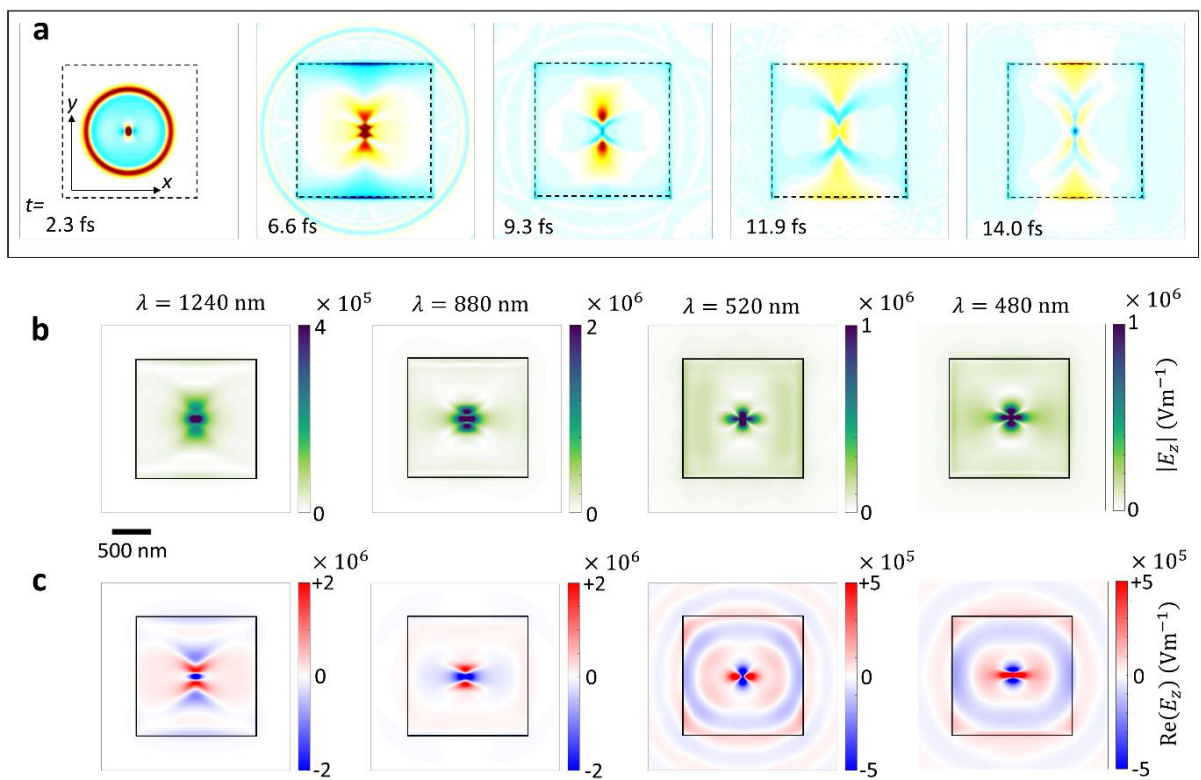

**Fig. 6| Simulated hyperbolic polaritons in a borophene flake.** (a) Snapshots of the simulated z-component of the electric field induced by a moving electron at the kinetic energy of 30 keV interacting with a borophene square flake with the edge length of 1600 nm, at depicted times after the arrival of the electron beam on top of the flake. (b) The absolute value and (c) the real value of the z-component of the Fourier-transformed electric field at depicted wavelengths.

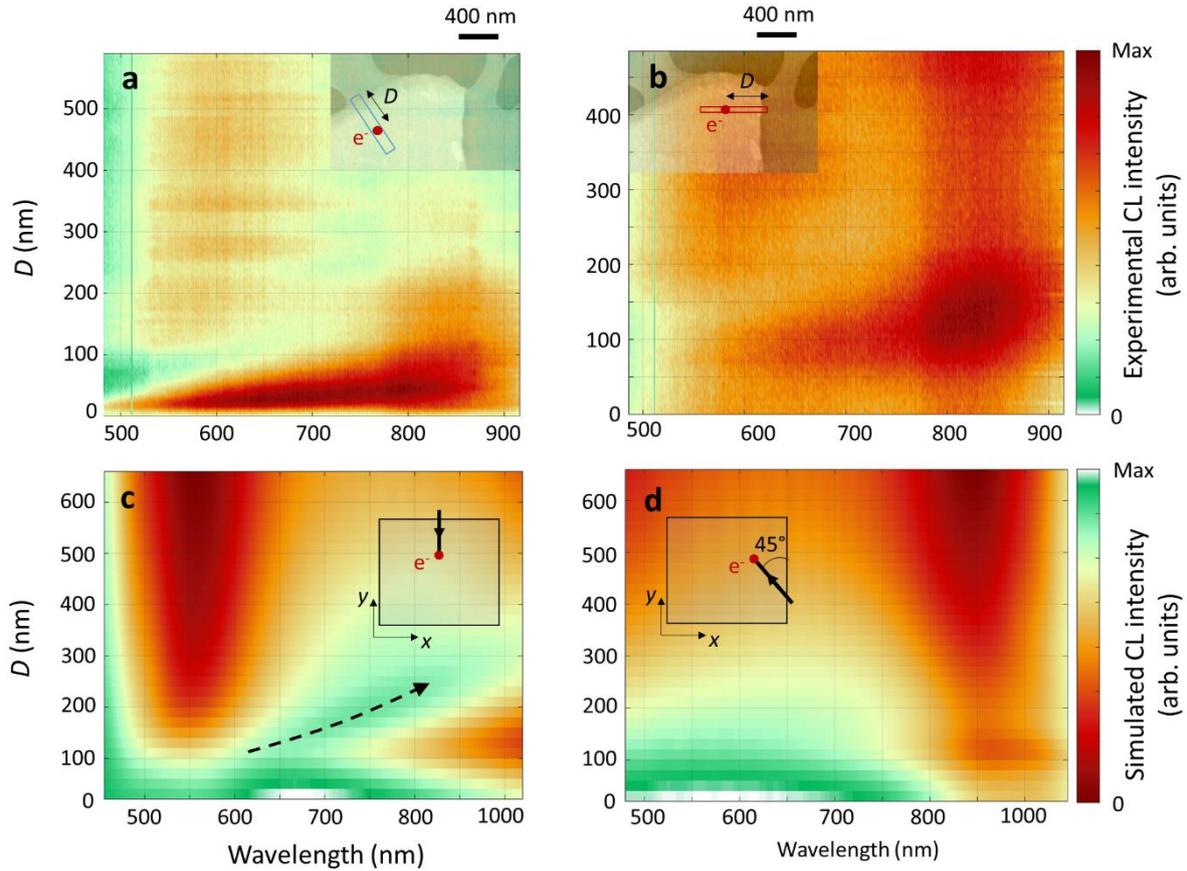

**Fig. 7 | Comparison between experimental and simulation results.** (a and b) Experimental wavelength-distance CL intensity maps, acquired along the scanning directions shown in insets. (c and d) Simulated CL maps for a square borophene flake with the edge wavelength of 1600 nm and thickness of 10 nm for depicted scanning directions.

To further confirm the differences in CL spectra acquired along different edges, we compare the experimental and simulated CL spectral-position maps. The experimental results are acquired from the same scanning directions provided in Fig. 5. A good comparison is obtained between the simulated and experimental maps. Particularly, the wavelength-dependent position of the minimum of the CL intensity with respect to the edge, shown by the dashed arrowed line is obvious in simulations as well. Slight differences between the simulation and experimental results are attributed to the different topologies of the simulated flake compared to the one used in experiments, particularly in the regions near to the edge. The interference fringes that are observed in the experimental map shown in Fig. 6a and 5b (right panels), at longer distances further away from the edge are attributed to artifacts, mainly caused by slight drifts while acquiring the data. This conclusion is confirmed by several facts: first, such interference fringes could not be observed in other flakes, second, the position of those fringes does not depend on the wavelength, and third, the simulated results do not exhibit such higher-order interferences.

Our results provide a first crucial insight into the physics of in-plane polaritons in borophene. We observe a strong in-plane anisotropic behavior, together with the formation of in-plane bound polariton waves. However, our calculations show a strong attenuation of the optical waves propagating along the y-axis, i.e. the direction along which the material behaves as a metal. The combination of this material with other 2D materials, such as hBN and borophene, could increase the propagation range of in-plane polaritons. In addition, the use of other techniques such as scanning tunneling microscopy to experimentally reveal the ultra-confined nature of polaritons in borophene will be a crucial next step. It is precisely because of the deep subwavelength confinement of borophene plasmons that borophene is a suitable platform for exploiting the nonlocal nature of polaritonic responses and exploring the light-matter interaction beyond the dipole approximation.

## Discussion

Here, we have explored the optical properties of synthesized free-standing borophene using deep-sub-wavelength cathodoluminescence spectroscopy. Using density functional theory, we have calculated the dielectric function along the principal axes and have shown that borophene is a material platform for the excitation of in-plane hyperbolic polaritons in the visible range. In-plane hyperbolic polaritons in borophene demonstrate an extreme subwavelength feature, with an effective wavelength that is 180 times smaller than the wavelength of the free-space light at visible photon energies. We have analytically calculated the dispersion of in-plane hyperbolic polaritons which features a highly directional propagation mechanism, and have further confirmed the excitation of hyperbolic polaritons using cathodoluminescence spectroscopy and polarimetry.

Extremely anisotropic hyperbolic polaritons have attracted tremendous attention recently, after the discovery of hyperbolic polaritons in hBN and a few other non-symmetric crystals, all happening at the far infrared range. Our results establish borophene as the only synthetic truly 2D material, that supports in-plane hyperbolic polaritons in the visible spectral range. This places borophene next to $MoOCl_2$ [78] the only materials supporting visible-range hyperbolic polaritons. Due to the impact of the hyperbolic response on tailoring the radiation from quantum emitters and manipulating the Purcell effect, borophene can be a material of choice for quantum-optical solid-state metrologies. In particular, when combined with deep-subwavelength characteristics of the in-plane polaritons in borophene, we anticipate that this 2D material can be used to manipulate the optical selection rules beyond the dipole approximation, at the visible range. Similarly, Dirac plasmons in graphene have been theoretically suggested for tailoring the optical selection rules; however, at the far-infrared range[82]. In addition, similar to graphene, this material could become an important candidate for light-driven electronic applications[83] and directional Schottky barriers in hybrid structures[84].

Exceptional optical properties and anisotropic metallic behavior of borophene, which are studied in this work, combined with its unique mechanical and electronic characteristics, position borophene as a promising candidate for a wide range of optical applications. The practical implications of borophene's optical properties are wide-ranging, including advanced optoelectronic devices to highly sensitive biosensors. These exceptional optical properties open up new ways for the development of innovative optoelectronic devices, such as photodetectors with enhanced responsivity and efficiency, which can be sensitive to light polarization. Additionally, borophene's ability to be integrated into flexible substrates makes it an ideal material for wearable technology and flexible optoelectronics. The practical implications of borophene's optical properties emphasize its potential to encourage various industries to use borophene in real-world scenarios. By investigating the unique characteristics of borophene and its integration into advanced technologies, we can better understand the groundbreaking impact of this remarkable 2D material.

## Methods

**Synthesis of borophene**

To grow borophene sheets, the experiment starts with cleaning a (100) silicon wafer using RCA-1 solution which is a mixture of 5 parts deionized water, including 1 part ammonium hydroxide (27%) and 1 part hydrogen peroxide (30%). Next, a 20 nm-thick layer of aluminum (Al) is deposited on the cleaned wafer by physical vapor deposition at a base pressure of $3 \times 10^{-6}$ Torr. The Al-coated silicon wafers are then placed in a chemical vapor deposition (CVD) chamber for the growth of borophene sheets. The CVD chamber is evacuated to a pressure of $3 \times 10^{-3}$ Torr, and the sample is then annealed for 1 hour at a temperature of 600 °C with a continuous flow of 20 sccm of $H_2$ gas.

The growth of borophene begins by introducing a gas mixture of 15 sccm $B_2H_6$ (diborane) and 40 sccm $H_2$ into the CVD chamber. After 10 minutes, the flow of diborane is stopped, and the sample is cooled down to room temperature over a period of 3 hours in an $H_2$ ambient atmosphere. The growth of 2D borophene sheets concludes with the sheets forming on the substrate, surrounded by aggregated Al islands (see Supplementary Fig. 2a and 2b). The growth mechanism has been previously reported in detail[51].

To avoid substrate effects on the plasmonic behavior of borophene and to remove Al aggregates, the as-grown borophene sheets are transferred onto the transmission electron microscopy (TEM) grids. The transfer process begins by spin-coating polymethyl methacrylate (PMMA) onto the borophene/silicon sample (Supplementary Fig.3). The sample is then heated at 360 K for 5 minutes to ensure good adhesion of the borophene sheets to the PMMA. After heating, the PMMA-coated sample is immersed in hydrochloric acid for a few seconds to remove the Al aggregates. The sample is then dipped into deionized (DI) water, and these two steps are repeated to separate the PMMA from the substrate and completely remove the Al aggregates. After repeating this process several times, the substrate is submerged in DI water while the PMMA/borophene layer remains floating, as shown in Supplementary Fig.3c. Finally, the layer is transferred on the copper grid (Supplementary Fig.3d) by dip coating, and the PMMA is removed by acetone (Supplementary Fig.3e). The borophene sheets are left isolated onto copper grid and ready for cathodoluminescence measurements (Supplementary Fig.3c). Thicker and square-like flakes are obtained by growing borophene on the $SiO_2$-coated Si wafers (Supplementary Fig.3d) which are suitable for investigating the polariton propagations.

**Cathodoluminescence**

For the CL measurements, the sample is carefully positioned under the aluminum parabolic mirror at the mirror's focal point to ensure optimal light collection and detection by the analyzer. This precise alignment allows photon counts reaching our CCD camera to exceed 10,000 per second. Parameters such as acquisition time, electron beam current, and beam spot size are determined through stepwise tests on the sample for each type of measurement, ensuring the avoidance of radiation damage. For spectral acquisition, the spot size was increased to 2 nm, scanning steps were adjusted from 5 nm to 20 nm, and the exposure time was reduced to 250 ms to prevent surface damage. The measured data are analyzed using the freeware ODEMIS, a Python-based script. The data are then exported to CSV files and imported into MATLAB for further analysis and visualization. Our CL detector (DELMIC SPARC system) and optical analyzers are integrated with a Zeiss-SIGMA field-emission scanning electron microscope (SEM).

**Structural and morphological characterization**

A Philips CM300 transmission electron microscopy is used to observe lattice images and SAED of borophene sheets at the acceleration voltage of 120kV. Zeiss-SIGMA field-emission SEM is used for morphological study of the samples. Raman spectroscopy is carried out using a Teksan-Opus Raman spectroscopy tool, utilizing Nd: YAG green laser (λ = 532 nm), low power of 7 mW, and a long-distance (LWD X60) objective lens was used to enhance the Raman signal and avoid destructing the borophene sheets.

**DFT calculations**

DFT calculations in our work are performed using Quantum ESPRESSO[85]. The electron–ion interaction is described with the optimized norm-conserving Vanderbilt pseudopotential[86]. The exchange and correlation energies are described using the generalized gradient approximation (GGA) based on the approach of Perdew-Burke-Ernzerhof (PBE)[87]. In all calculations, the plane wave cut-off energy is 50 Hartree. The relaxation was carried out with an $8 \times 8 \times 1$ k-point mesh and a force threshold of $10^{-5}$. We performed the calculation of the dielectric tensor through the following sequential steps: first, we performed the initial self-consistent field (SCF) calculation by employing an $8 \times 8 \times 1$ k-point mesh and including 16 bands. Subsequently, we conducted a more refined calculation using a finer $32 \times 32 \times 1$ k-point mesh and 1000 bands for the non-self-consistent field (NSCF) calculation. Finally, a post-processing (PP) step was executed to obtain the dielectric tensor.

## Data availability

All data supporting the findings of this study are available within the article and the Supporting Information file, or available from the corresponding authors upon request.

## Acknowledgements

Authors acknowledge fruitful discussions with Dr. Davoodi (Kiel University). This project has received funding from the European Research Council (ERC) under the European Union's

Horizon 2020 research and innovation programme under grant agreement no. 802130 (Kiel, NanoBeam) and grant agreement no. 101017720 (EBEAM), and Deutsche Forschungsgemeinschaft.

Supporting Information

**Two-dimensional borophene: In-plane hyperbolic polaritons in the visible spectral range**

*Yaser Abdi[1,2]\*, Masoud Taleb[2], Soheil Hajibaba[1], Mohsen Moayedi[1], Nahid Talebi[2]\**

1. Department of Physics, University of Tehran, 1439955961 Tehran, Iran

2. Institute of Experimental and Applied Physics, Kiel University, 24118 Kiel, Germany
   \* E-Mail: y.abdi@ut.ac.ir,  talebi@physik.uni-kiel.de

**Content:**
1. **Band structure and density of state of $\chi_3$ borophene**
2. **In-plane polaritons in an atomically thin borophene sheet**
3. **Borophene's synthesis and morphological images**
4. **Modelling borophene permittivity for time-domain simulations**

1. **Band structure and density of state of $\chi_3$ borophene.**

The electronic band structure of $\chi_3$ borophene, obtained through DFT calculations using Quantum ESPRESSO, is shown in Supplementary Fig. 1a. Energy levels are plotted along a G-M-X-G-N-Y-G path in the Brillouin Zone's reciprocal space. This band structure highlights the anisotropic nature of $\chi_3$ borophene. In certain directions, such as X-G, the electron energy bands intersect the Fermi level, indicating metallic behavior. Conversely, in the X-M direction, the electron bands near the Fermi level are relatively flat, suggesting more localized electronic states and directional variation in electronic properties. Additionally, $\chi_3$ borophene has a nonzero density of states (DOS) at the Fermi level (Supplementary Fig. 1b). As a result, $\chi_3$ borophene exhibits metallic behavior with pronounced anisotropy, and its electrical conductivity is confined along the atomic strips parallel to the y direction.

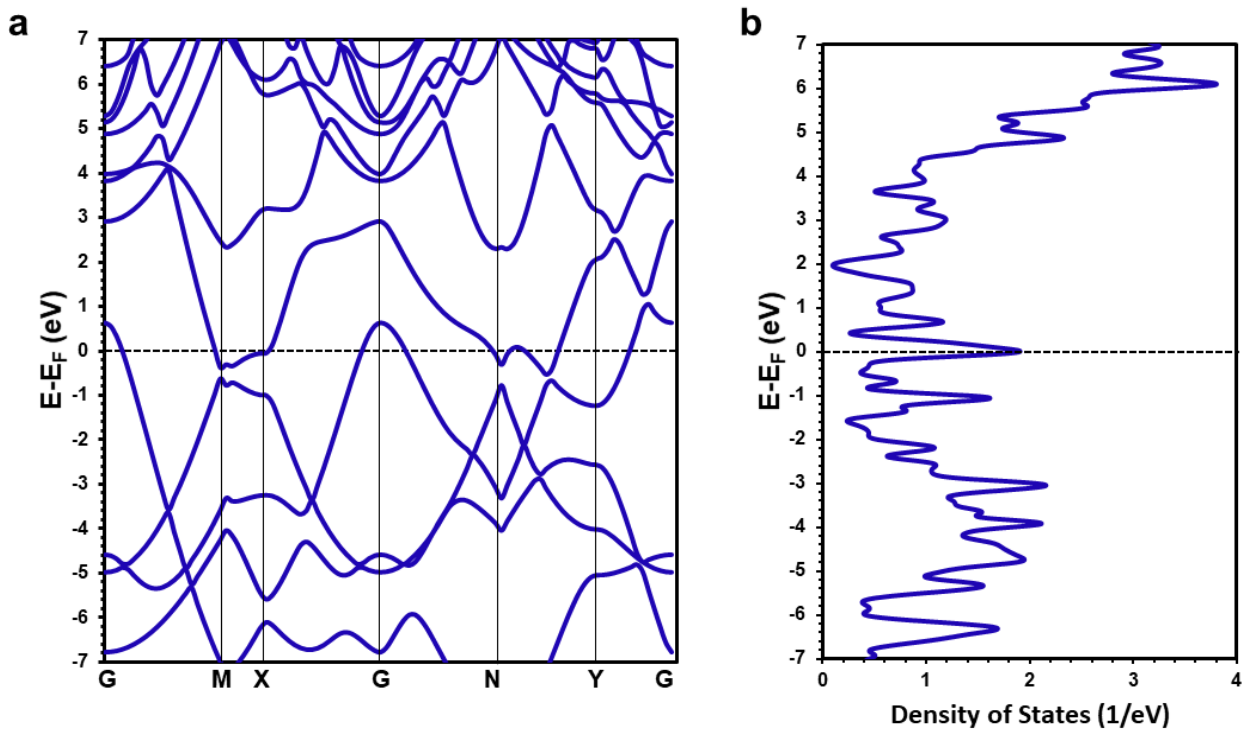

**Supplementary Fig. 1| Electronic behavior of $\chi_3$ borophene obtained by DFT calculation.** (a) Band structure along the G-M-X-G-N-Y-G path in the reciprocal space of the Brillouin Zone. (b) Density of state showing the nonzero states near the Fermi level.

## 2. In-plane polaritons in an atomically thin borophene sheet

We consider a borophene sheet located at the z=0 plane, supporting in-plane surface waves confined to the plane and propagating along the direction specified by the azimuthal angle $\varphi$ with respect to the principal axis x. We consider the vector potential approach allowing us to decompose the waves into transverse-magnetic (TM) and transverse-electric (TE) parts, by choosing $\vec{A} = (0,0,A_z)$ and $\vec{F} = (0,0,F_z)$, with $\vec{A}$ and $\vec{F}$ being the magnetic and electric vector potentials. Therefore, the z-components of the magnetic and electric vector potentials are constructed as

$$A_z(x,y,z) = \begin{cases} A_1 e^{-\kappa_z z} e^{ik_\| \cos\varphi x + ik_\| \sin\varphi y} & z > 0 \\ A_2 e^{+\kappa_z z} e^{ik_\| \cos\varphi x + ik_\| \sin\varphi y} & z < 0 \end{cases} \quad (1)$$

and

$$F_z(x,y,z) = \begin{cases} F_1 e^{-\kappa_z z} e^{ik_\| \cos\varphi x + ik_\| \sin\varphi y} & z > 0 \\ F_2 e^{+\kappa_z z} e^{ik_\| \cos\varphi x + ik_\| \sin\varphi y} & z < 0 \end{cases} \quad (2)$$

respectively. The electric and magnetic field components are obtained as

$$\vec{E}(x,y,z) = -\vec{\nabla} \times \vec{F} - \frac{1}{i\omega\varepsilon_0} \vec{\nabla} \times \vec{\nabla} \times \vec{A} \quad (3)$$

and

$$\vec{H}(x,y,z) = \vec{\nabla} \times \vec{A} - \frac{1}{i\omega\mu_0} \vec{\nabla} \times \vec{\nabla} \times \vec{A}, \quad (4)$$

Respectively. Using Supplementary Equations (1) to (4) and applying the boundary conditions as

$\hat{z} \times (\vec{H}_1 - \vec{H}_2) = \hat{\sigma} : \vec{E}_\|$ and $\hat{z} \times (\vec{E}_1 - \vec{E}_2) = 0$, one obtains $A_1 = -A_2$ and $F_1 = F_2$, and the following relation between the $A_i$ and $F_i$ coefficients:

$$\left(+2 - \sigma_{yy} \frac{\kappa_z}{i\omega\varepsilon_0}\right) k_\| \sin\varphi A_1 + \left(2\frac{\kappa_z}{i\omega\mu_0} - \sigma_{yy}\right) k_\| \cos\varphi F_1 = 0 \quad (5)$$

and

$$\left(-2 + \sigma_{xx} \frac{\kappa_z}{i\omega\varepsilon_0}\right) k_\| \cos\varphi A_1 + \left(2\frac{\kappa_z}{i\omega\mu_0} - \sigma_{xx}\right) k_\| \sin\varphi F_1 = 0 \quad (6)$$

Simultaneous satisfaction of Supplementary Equations (5) and (6) leads to the following values for the decay ratio $\kappa_z$:

$$\kappa_z = \frac{k_0^2}{4i\omega(\sigma_{yy}\sin^2\varphi + \sigma_{xx}\cos^2\varphi)} \left\{ -\left(\frac{4}{\mu_0} + \frac{\sigma_{xx}\sigma_{yy}}{\varepsilon_0}\right) \pm \sqrt{\left(\frac{4}{\mu_0} + \frac{\sigma_{xx}\sigma_{yy}}{\varepsilon_0}\right)^2 - 16c^2(\sigma_{yy}\sin^2\varphi + \sigma_{xx}\cos^2\varphi)(\sigma_{xx}\sin^2\varphi + \sigma_{yy}\cos^2\varphi)} \right\}$$

as shown in the main text, as well as the following results for the in-plane components of the electric field:

$$E_\alpha = +A_1 \left\{ \frac{\kappa_z}{i\omega\mu_0} + \left(2 - \sigma_{\alpha\alpha}\frac{\kappa_z}{i\omega\mu_0}\right)\left(2\frac{\kappa_z}{i\omega\varepsilon_0} - \sigma_{\alpha\alpha}\right)^{-1} \right\} (ik_\alpha) e^{-\kappa_z |z|} e^{i\vec{k}_\| \cdot \vec{r}_\|} \quad (7)$$

and the tangential components of the magnetic field represented as $\vec{H}_\| = \frac{1}{\eta_0} \hat{z} \times \vec{E}_\|$.

## 3. Borophene's synthesis and morphological images

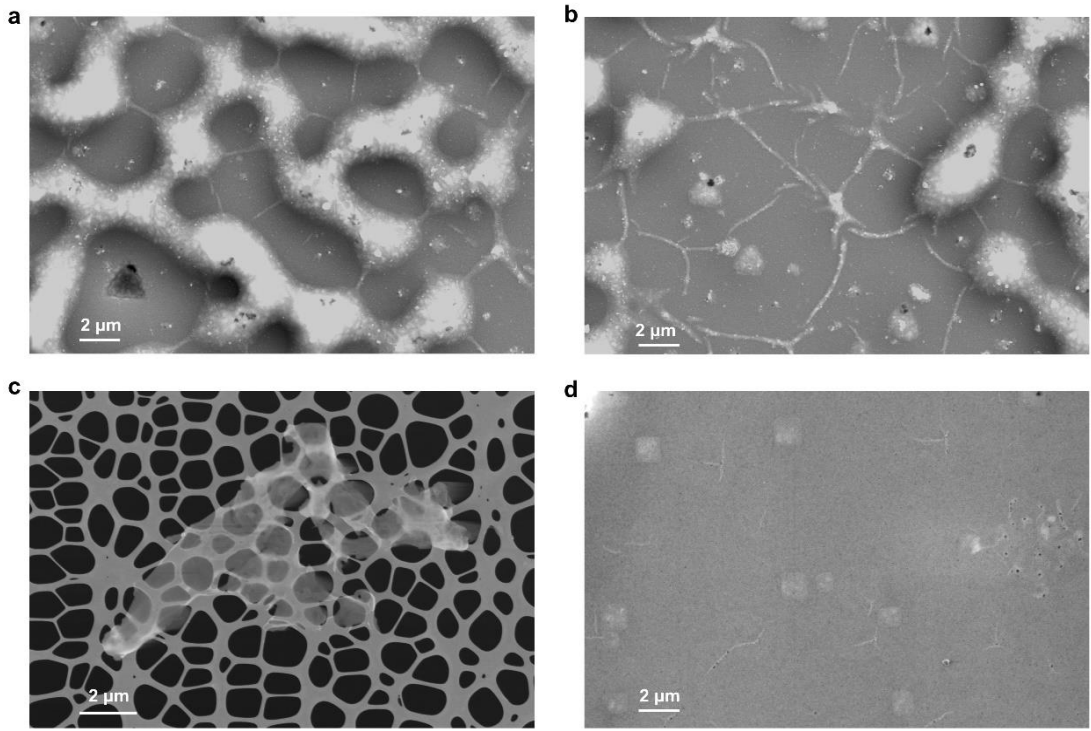

**Supplementary Fig. 2| Scanning electron microscope images**. **(a)** As-grown borophene sheets between aluminum islands on a Si wafer. **(b)** Large-area borophene grown on Si with small aluminum aggregates. **(c)** Borophene sheets transferred onto holey carbon after aluminum removal. **(d)** Square-like borophene obtained by growth on a $SiO_2$-coated wafer.

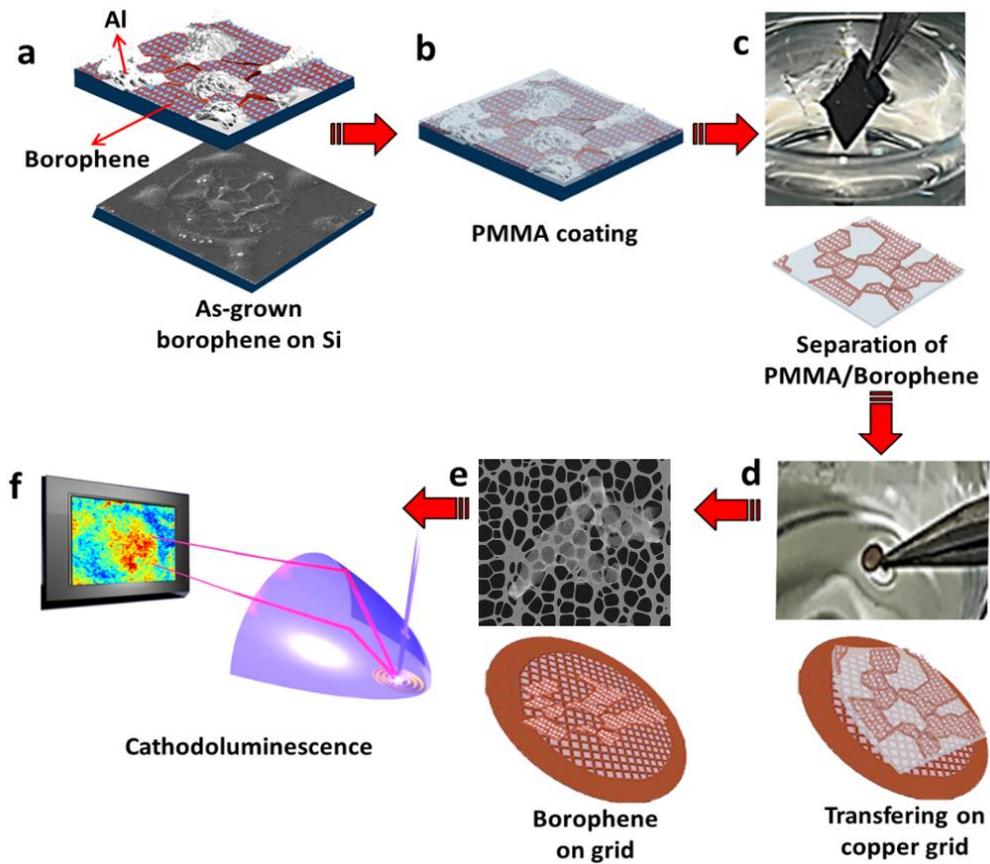

**Supplementary Fig. 3| Schematical illustration of the synthesis process**. **(a)** The bottom image shows a SEM image of as-grown borophene sheets on a silicon substrate. The top image shows the schematic of the same sample with white islands (bright area shown in SEM) representing Al aggregations and the grown sheets between the Al islands representing borophene. **(b)** PMMA (Poly methyl methacrylate) is coated on the sample. **(c)** PMMA-coated borophene is separated from the substrate and Al aggregates are removed using the transfer method explained in the text. **(d)** Floated PMMA/borophene is transferred onto a copper grid by fishing. **(e)** SEM (top) and schematic (bottom) images of borophene coated grid after removing the PMMA. **(f)** The schematic shows the cathodoluminescence setup.

### 4. Modelling borophene permittivity for time-domain simulations

For modelling the permittivity of borophene for finite-difference time-domain simulations, we use Drude model in addition to two critical point functions[88]. This model is described as

$$\varepsilon_{r\alpha\alpha} = \varepsilon_{r\alpha\alpha}^{\infty} - \frac{1}{\lambda_{p,\alpha}^2 \left(1/\lambda^2 + i/\gamma_{p,\alpha}\lambda\right)} + \sum_{n=1}^{2} \frac{A_{\alpha,n}}{\lambda_{\alpha,n}} \left[ \frac{e^{i\varphi_{\alpha,n}}}{1/\lambda_{\alpha,n} - 1/\lambda - i/\gamma_{\alpha,n}} + \frac{e^{-i\varphi_{\alpha,n}}}{1/\lambda_{\alpha,n} + 1/\lambda + i/\gamma_{\alpha,n}} \right] \quad (8)$$

where the first term is the Drude model and the second term includes two-critical point functions. Fitting the borophene permittivity obtained using DFT calculations with this model, we obtain the parameters outlined in Supplementary Table 1.

**Supplementary Table 1|** Fitting parameters of the Drude model+ two-critical point functions for borophene.

|  | $\varepsilon_{r\alpha\alpha}^{\infty}$ | $\lambda_{p,\alpha}$ (nm) | $\gamma_{p,\alpha}$ (nm) | $A_{\alpha,1}$ | $\lambda_{\alpha,1}$ (nm) | $\gamma_{\alpha,1}$ | $\varphi_{\alpha,1}$ | $A_{\alpha,2}$ | $\lambda_{\alpha,2}$ (nm) | $\gamma_{\alpha,2}$ (nm) | $\varphi_{\alpha,2}$ |
|---|---|---|---|---|---|---|---|---|---|---|---|
| $\varepsilon_{r,xx}$ | 1 | 2318 | 42000 | 1 | 434.02 | 1677 | -25° | 1 | 925.24 | 2377 | -55° |
| $\varepsilon_{r,yy}$ | 1 | 568 | 4500 | 1.1 | 1509.02 | 12513 | -26° | 1.2 | 2200.25 | 11529 | -5.4° |